\documentstyle[preprint,tighten,aps,amstex,graphicx,times,floats]{revtex}

\newcommand{\eg}{{\it e.g.}\;}
\newcommand{\ie}{{\it i.e.}\;}

\newcommand{\ibid}{{\it ibid.}\;}

\newcommand{\msbar}{\overline{{\rm MS}}}

\newcommand{\SP}{\scriptscriptstyle}
\newcommand{\nni}{\tilde{\chi}_i^0}
\newcommand{\nnj}{\tilde{\chi}_j^0}
\newcommand{\nne}{\tilde{\chi}_1^0}
\newcommand{\nnz}{\tilde{\chi}_2^0}

\newcommand{\nnv}{\tilde{\chi}_4^0}
\newcommand{\cpi}{\tilde{\chi}_i^+}

\newcommand{\cpe}{\tilde{\chi}_1^+}

\newcommand{\cpme}{\tilde{\chi}_1^\pm}

\newcommand{\cpmi}{\tilde{\chi}_i^\pm}
\newcommand{\slmrl}{\tilde{\ell}^{^-}_{\SP L,R}}
\newcommand{\snl}{\tilde{\nu}_{\SP L}}
\newcommand{\snlb}{\bar{\tilde{\nu}}_{\SP L}}
\newcommand{\snlsnlb}{\hspace*{-1ex}
                      \stackrel{\,\mbox{\tiny (}{\SP -}\mbox{\tiny )}}
                               {\tilde{\nu}}_{\hspace*{-0.5ex}\SP L}}
\newcommand{\sepl}{\tilde{e}_{\SP L}^+}
\newcommand{\seml}{\tilde{e}_{\SP L}^-}
\newcommand{\sepr}{\tilde{e}_{\SP R}^+}
\newcommand{\semr}{\tilde{e}_{\SP R}^-}
\newcommand{\sepml}{\tilde{e}_{\SP L}^\pm}

\newcommand{\nlnlb}{\!\stackrel{\!\mbox{\tiny (}{\SP -}\mbox{\tiny )}}
                               {\nu_{\ell}}}
\newcommand{\qbqbp}{\bar{q}^{\,\SP (\hspace{0.7ex})}\hspace*{-1.15ex}'
                    \hspace*{0.5ex}}

\newcommand{\epc}[3]{${\rm Eur. Phys. J.}$ {\bf C#1} (19#2) #3}
\newcommand{\zpc}[3]{${\rm Z. Phys.}$ {\bf C#1} (19#2) #3}
\newcommand{\npb}[3]{${\rm Nucl. Phys.}$ {\bf B#1} (19#2)~#3}
\newcommand{\plb}[3]{${\rm Phys. Lett.}$ {\bf B#1} (19#2) #3}
\newcommand{\plbold}[3]{${\rm Phys. Lett.}$ {\bf #1B} (19#2) #3}
\renewcommand{\prd}[3]{${\rm Phys. Rev.}$ {\bf D#1} (19#2) #3}
\renewcommand{\prl}[3]{${\rm Phys. Rev. Lett.}$ {\bf #1} (19#2) #3}
\newcommand{\prep}[3]{${\rm Phys. Rep.}$ {\bf #1} (19#2) #3}

\newcommand{\mpl}[3]{${\rm Mod. Phys. Lett.}$ {\bf #1} (19#2) #3}

\newcommand{\hep}[1]{${\tt hep\!-\!ph/}${#1}}
\newcommand{\hex}[1]{${\tt hep\!-\!ex/}${#1}}

\begin{document}

\thispagestyle{empty}

\title{ 
~\vspace*{-2cm} \\
{\normalsize \rm
\begin{flushright}
CERN--TH/99--159 \\
DESY 99--055 \\
DTP/99/44 \\
MAD--PH--99--1114 \\
ANL-HEP-PR-99-71 \\
hep-ph/9906298 \\
\end{flushright} }
The Production of Charginos/Neutralinos and Sleptons at 
Hadron Colliders
} 

\author{
W.~Beenakker${}^1$\footnote{Supported by a PPARC Research Fellowship},
M.~Klasen${}^2$\footnote{Supported by DOE grant W-31-109-ENG-38},
M.~Kr\"amer${}^3$\footnote{Supported in part by the EU FF Programme 
                           under contract FMRX-CT98-0194 (DG 12 - MIHT)}, 
T.~Plehn${}^4$\footnote{Supported in part by DOE grant DE-FG02-95ER-40896 
                       and in part by the University of Wisconsin Research 
                       Committee with funds granted by the Wisconsin Alumni 
                       Research Foundation}, 
M.~Spira${}^5$\footnote{Heisenberg Fellow}, and
P.M.~Zerwas${}^6$
} 

\address{\vspace*{2ex} 
${}^1$ Department of Physics, 
       University of Durham, Durham DH1 3LE, U.K.\\
${}^2$ Argonne National Laboratory, Argonne, IL 60439, USA \\
${}^3$ Theoretical Physics Division, CERN, CH-1211 Geneva 23, Switzerland \\
${}^4$ Department of Physics, 
       University of Wisconsin, Madison, WI 53706, USA \\
${}^5$ II. Institut f\"ur Theoretische Physik$^\ddag$,
       Universit\"at Hamburg, D-22603 Hamburg, FRG \\
${}^6$ Deutsches Elektronen--Synchrotron DESY, 
       D-22603 Hamburg, FRG \\
} 

\maketitle 

\begin{abstract} 
We analyse the production of charginos/neutralinos and sleptons at the
hadron colliders Tevatron and LHC in the direct channels: $p\bar{p}/pp
\to \tilde{\chi}_i \tilde{\chi}_j + X$ and $\tilde{\ell}
\bar{\tilde{\ell}'} +X$. The cross sections for these reactions are
given in next-to-leading order SUSY QCD. By including the higher-order
corrections, the predictions become theoretically stable, being nearly
independent of the factorization and renormalization scales.  Since
the corrections increase the cross sections, the discovery range for
these particles is extended in the refined analysis.
\end{abstract}

%\pacs{PACS numbers: 13.85.-t, 14.80.Bn, 14.60.Fg} 
% these PACS were chosen for ph1057 %

\vspace*{5mm}

%%%%%%%%%%%%%%%%%%%%%%%%%%%%%%% MAIN TEXT %%%%%%%%%%%%%%%%%%%%%%%%%%%%
Non-coloured supersymmetric particles, \ie char\-ginos, neutra\-linos
and sleptons, can be searched for at hadron colliders in cascade
decays of squarks/gluinos and in the direct production
channels~\cite{born}
\begin{equation} 
  p\bar{p}/pp \to \tilde{\chi}_i \tilde{\chi}_j + X
  \qquad \mbox{and} \qquad
  p\bar{p}/pp \to \tilde{\ell}\bar{\tilde{\ell}'} + X.
\end{equation} 
In the minimal supersymmetric extension of the Standard Model (MSSM)
the two charginos $\tilde{\chi}_{1,2}^\pm$ are mixtures of 
charged winos and
higgsinos, while the four neutralinos $\nne,\ldots,\nnv$ are
mixtures of the neutral wino, the bino, and the two neutral
higgsinos~\cite{mssm}.  The scalar partners of the chiral lepton
states are denoted by $\tilde{\ell}, \tilde{\ell}' =
\slmrl, \snl$. 

When the LEP2 running will be terminated, with either positive or
negative results, and before prospective lepton colliders will start
operation, the hadron colliders Tevatron and LHC will be the only
machines to search for supersymmetric particles. In order to exploit
these machines exhaustively, a proper understanding of the
hadroproduction mechanisms of supersymmetric particles is mandatory.
On the theoretical side this demands the control of higher-order
SUSY-QCD corrections. They reduce the (artificial) dependence of the
cross sections on the renormalization and factorization scales in
leading order and refine the numerical accuracy of the theoretical
predictions. This program has been carried out for the coloured gluinos
and squarks~\cite{sq/gl,sq_gl}, including a special analysis for top
squarks~\cite{stop}. The associated production of gluinos with
charginos/neutralinos has been addressed recently~\cite{asso}.  QCD
corrections to the Drell--Yan production of slepton pairs have been
analysed in Ref.~\cite{bhr}. In this letter we present the complete
SUSY-QCD analyses for the production of all possible pairs of
non-coloured supersymmetric particles.\medskip

\noindent {\bf 1.} In many models, the non-coloured
charginos, neutralinos and sleptons belong to the class of the lightest
supersymmetric particles. The low masses (partly) counterbalance the
small cross sections for direct production. Moreover, the classical
reaction $p\bar{p}/pp \to \cpme \nnz$ with subsequent decays $\cpme
\to \nne \ell^\pm \nlnlb$ and $\nnz \to \nne \ell^+ \ell^-$ leads to
gold-plated $\ell^\pm \ell^+ \ell^-$ trilepton
signatures~\cite{trilepton0}.  Such trilepton signatures have been
exploited in several CDF and D0 analyses at the
Tevatron~\cite{trilepton}, leading to bounds in the MSSM parameter
space that are similar to those from LEP2. Increasing energies and
luminosities will significantly extend the range of sensitivity in the
near future, see \eg Refs.~\cite{kane_book,atlas}.

\begin{figure}[htb] 
\begin{center}
\includegraphics[width=10.0cm]{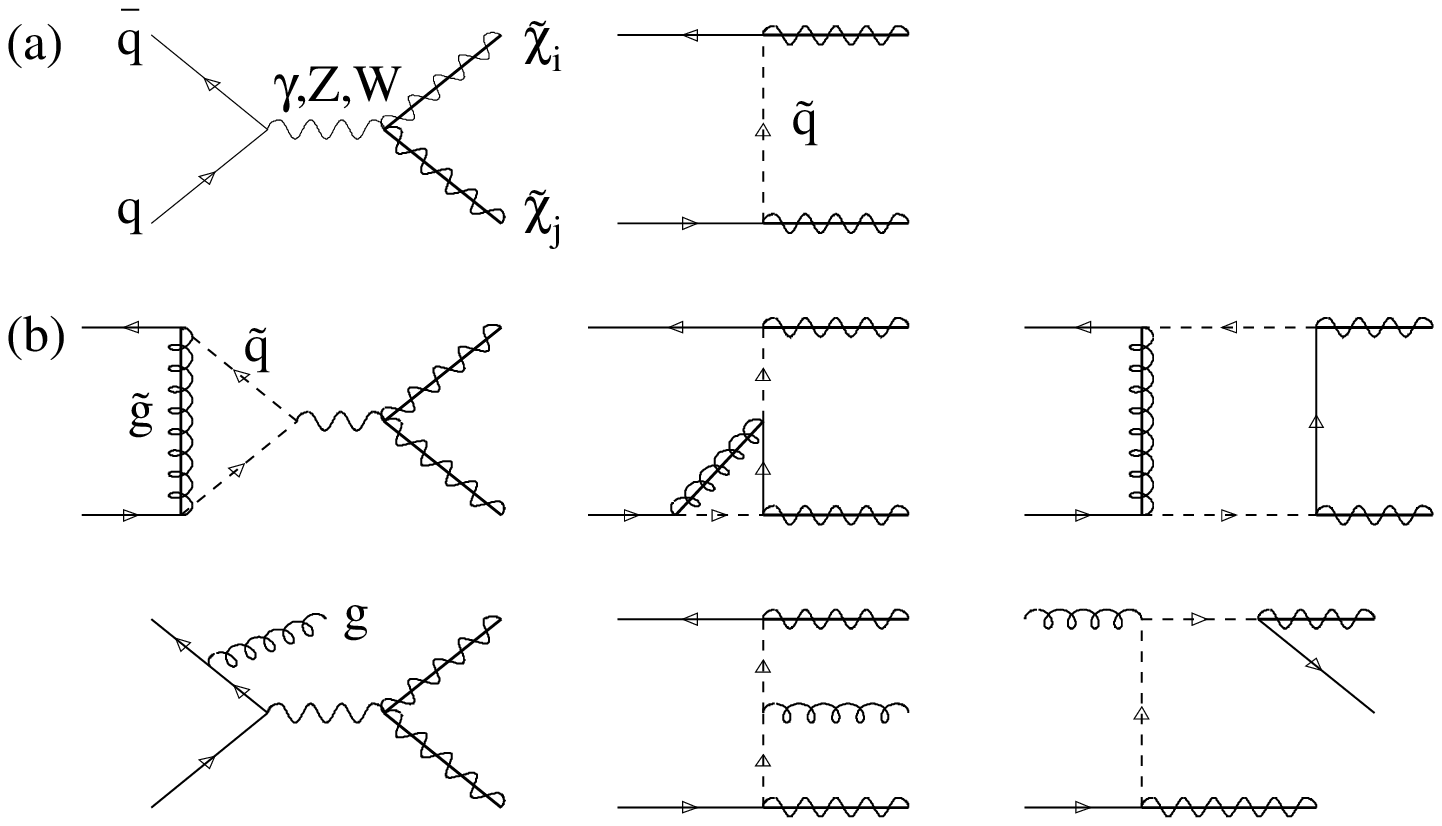}
\caption[]{\label{fg:feyn} 
    Basic diagrams for the production of chargino/neutralino pairs
    at hadron colliders in quark-antiquark collisions; (b) generic
    diagrams of SUSY-QCD corrections, comprising vertices, box diagrams, 
    and diagrams for three-parton final states.    
}
%\end{center} 
%\end{figure}
%\begin{figure}[htb] 
%\begin{center}
\vspace*{3cm}
\includegraphics[width=9.0cm]{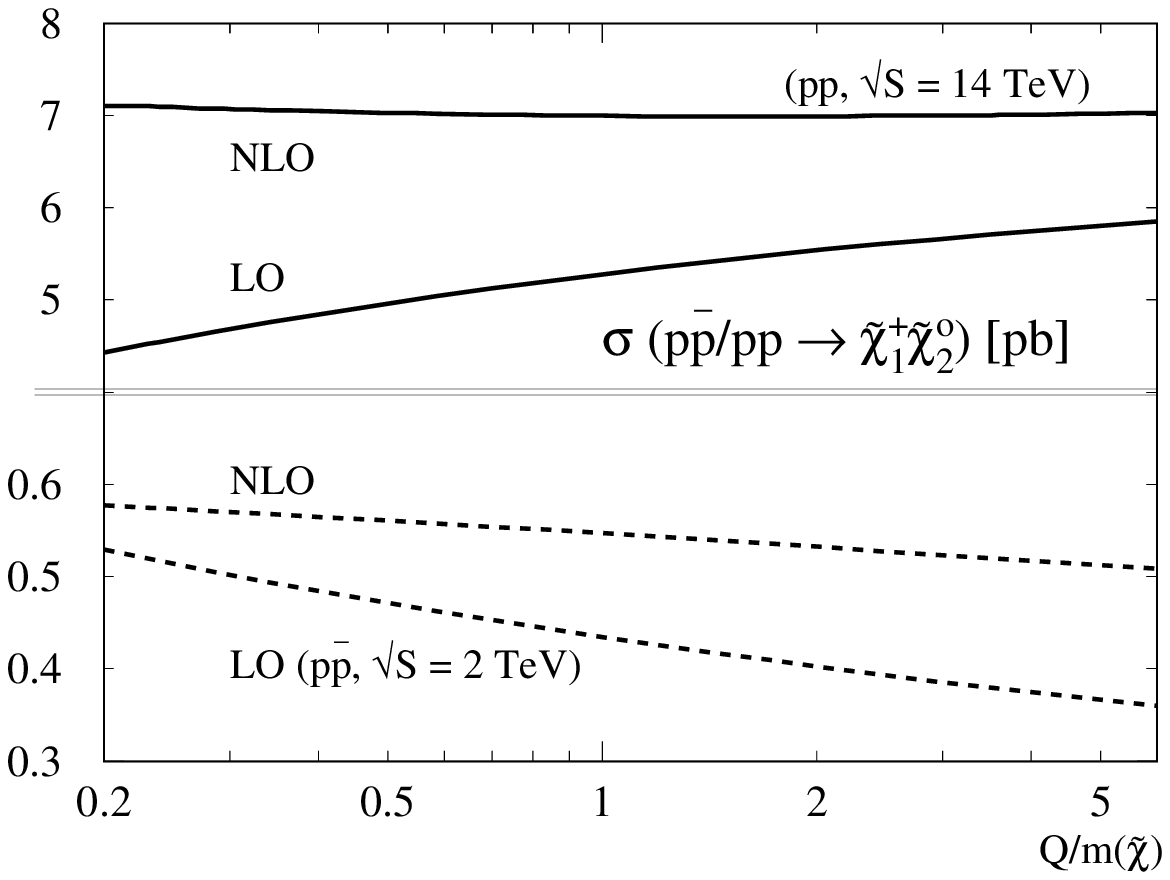}
\caption[]{\label{fg:scale_neut} 
    Dependence of the cross sections for the production of $\cpe \nnz$
    pairs on the factorization/renormalization scale $Q$ in leading
    order and next-to-leading order SUSY QCD for the mSUGRA point
    defined in the text. The scale $Q$ is given in
    units of the average of the $\cpe$ and $\nnz$ masses $m(\tilde{\chi})$. 
}
\end{center} 
\end{figure}

The basic diagrams for the production of pairs of charginos and
neutralinos are depicted in Fig.~\ref{fg:feyn}(a) at the parton
level. The vector bosons ($V\!=\!\gamma/Z/W$) in the $s$-channel
couple to the gaugino ($\lambda$) and higgsino ($\psi_H$) components
of the charginos and neutralinos, whereas the $u/t$-channel squark
diagrams, in the limit where the light-quark masses are neglected, 
involve only gaugino components. In terms of two-component Weyl
spinors, the chargino/neutralino mass eigenstates read
\begin{alignat}{6}
&\chi_i^+  = V_{ij}\,\psi_j^+ \quad & ,\quad \text{basis:} \quad 
& \psi_j^+ = \left( -i \lambda^+, \psi_{H_2}^+ \right) \notag \\
&\chi_i^-  = U_{ij}\,\psi_j^- \quad & ,\quad \text{basis:} \quad 
& \psi_j^- = \left( -i \lambda^-, \psi_{H_1}^- \right) \notag \\
&\chi_i^0  = N_{ij}\,\psi_j^0 \quad & ,\quad \text{basis:} \quad 
& \psi_j^0 = \left( -i \lambda', -i \lambda^3, 
                             \psi_{H_1}^0, \psi_{H_2}^0 \right).
\end{alignat}
By combining the two-component Weyl spinors, the four-component Dirac
spinors $\cpmi$ and Majorana spinors $\nni$ can be constructed.  The
mixing matrices $U,V$ and $N$ are defined such that the chargino and
neutralino masses are real and positive~\cite{mssm,haber_gunion}.
After Fierz transformations of the squark-exchange amplitudes, the
transition matrix element can be expressed in terms of four bilinear
charges $Q_{\alpha\beta}$~\cite{lc}, coefficients of the associated
quark and gaugino currents carrying chiralities $\alpha,\beta=L,R$.
For example, the partonic process $q\bar q' \to \cpi \nnj$, upon
neglecting generational mixing in the quark and squark sectors, is
described by the bilinear charges:
\begin{alignat}{6}
& Q_{LL} = 
  \frac{1}{\sqrt{2}\,s_W^2}\left[ 
     \frac{N_{j2}^*\,V_{i1} - N_{j4}^*\,V_{i2}/\sqrt{2}}{s-M_W^2}
    +\frac{V_{i1}}{c_W}\,
     \frac{N_{j1}^*\,(e_{\tilde{q}}-I_{3\tilde{q}})\,s_W
          + N_{j2}^*\,I_{3\tilde{q}}\,c_W}{u-m^2_{\tilde{q}}}
                         \right]
\notag \\
& Q_{LR} = 
  \frac{1}{\sqrt{2}\,s_W^2}\left[
     \frac{N_{j2}\,U_{i1}^* + N_{j3}\,U_{i2}^*/\sqrt{2}}{s-M_W^2}
    -\frac{U_{i1}^*}{c_W}\,
     \frac{N_{j1}\,(e_{\tilde{q}'}-I_{3\tilde{q}'})\,s_W
          + N_{j2}\,I_{3\tilde{q}'}\,c_W}{t-m^2_{\tilde{q}'}}
                         \right]
\notag \\[1ex]
& Q_{RL} = Q_{RR} = 0,
\end{alignat}
with $s=(p_{q}+p_{\bar{q}'})^2$, $t=(p_{q}-p_{\tilde{\chi}_i})^2$ and
$u=(p_{q}-p_{\tilde{\chi}_j})^2$. The electric charges and third
isospin components of the exchanged squarks are denoted by
$e_{\tilde{q}}$ and $I_{3\tilde{q}}$, respectively.  Furthermore we
define the cosine $c_W$ and the sine $s_W$ of the weak mixing angle by
$s_W^2=1-M_W^2/M_Z^2$.  The generic form of the leading-order partonic
cross section after spin and colour averaging reads
\begin{alignat}{6}
& \frac{d\hat{\sigma}}{d t}[q\qbqbp\to\tilde{\chi}_i\tilde{\chi}_j] 
= \frac{\pi\alpha^2}{3s^2}
  \left[ (|Q_{LL}|^2+|Q_{RR}|^2)\, u_i u_j 
       + (|Q_{LR}|^2+|Q_{RL}|^2)\, t_i t_j \right.
\notag \\
&
\left. \hspace*{4.2cm} {}+ 
2\,\mbox{Re}(Q_{LL}^{*}Q_{LR}+Q_{RR}^{*}Q_{RL})\,m_{\tilde{\chi}_i}
m_{\tilde{\chi}_j}s
\vphantom{|Q_{LL}|^2} \right], 
\end{alignat}
with the abbreviations $\,t_{i,j}=t-m^{2}_{\tilde{\chi}_{i,j}}$
and $u_{i,j}=u-m^{2}_{\tilde{\chi}_{i,j}}$.\smallskip

SUSY-QCD corrections involve quark/gluon and squark/gluino
diagrams. Leaving aside the standard QCD diagrams, generic SUSY-QCD
diagrams for $qqV$ and $q\tilde{q}\tilde{\chi}$ vertex corrections,
box diagrams and diagrams of three-parton final states are shown in
Fig.~\ref{fg:feyn}(b). The virtual corrections have been evaluated in
the $\msbar$ renormalization scheme, with the $\tilde{q}$ and
$\tilde{g}$ masses defined on-shell. The artificial breaking of
supersymmetry by the mismatch of $2$ gaugino and $(D-2)$ transverse
vector degrees of freedom in $D \ne 4$ dimensions is compensated by
finite counter terms~\cite{sq_gl,martin_vaughn}. In this way, the
supersymmetry can be restored by modifying the bare Yukawa
$q\tilde{q}\tilde{\chi}$ coupling $\hat{g}$ with respect to the
associated gauge coupling $g$: $\hat{g}=g [1-\alpha_s/(6\pi)]$. The
infrared and collinear singularities of the three-parton cross
sections are extracted by applying the dipole subtraction
method~\cite{dipole}. The virtual and real corrections are different
for the $s$-channel and $t/u$-channel exchange mechanisms of the gauge
bosons and squarks. This affects strongly any destructive interference
effects that may be present between leading-order diagrams and may
thus give rise to large $K$-factors.

The inelastic Compton process in Fig.~\ref{fg:feyn}(b) evolves through
a squark state, which can decay as an on-shell state into $\,q
\tilde{\chi}\,$ if $\,m_{\tilde{q}}>m_{\tilde{\chi}}$. To avoid double
counting, this resonance contribution is removed from the continuum
$p\bar{p}/pp \to \tilde{\chi} \tilde{\chi}$ ensemble; it is counted
naturally in the mixed $\tilde{q} \tilde{\chi}$ ensemble. The
separation is technically defined by subtracting the resonance part
$\sigma^{\rm Res} = \hat{\sigma}[qg\to\tilde{q}\tilde{\chi}]\, {\rm
BR}[\tilde{q}\to\tilde{\chi}q]$ in the narrow-width approximation for
the squark state~\cite{sq_gl}.\smallskip

The QCD corrections to the cross sections will be illustrated in the
mSUGRA scenario for the specific CP-conserving point 
[$m_{1/2}=150$~GeV, 
$m_0=100$~GeV,
$A_0=300$~GeV,
$\mu>0$,  
$\tan\beta=4$].
The parameters $m_{1/2}$ and $m_0$ are the universal gaugino and
scalar masses at the GUT scale, and $A_0$ is the universal trilinear
coupling in the superpotential. From these five parameters, all the
masses and couplings are determined by the evolution from the GUT
scale down to the low electroweak scale~\cite{drees_martin}, leading
for the mSUGRA point defined above to the masses: $\cpme/ \nne/ \nnz =
101/56/104$~GeV, $\tilde{q}/\tilde{g} = 359/406$~GeV, and $\mu/M_1/M_2
= 278/62/123$~GeV. In this specific example the light charginos and
neutralinos are gaugino-like. Chargino/neutralino masses are varied by
varying $m_{1/2}$ while leaving the other parameters
unchanged.\smallskip

\begin{figure}[htb] 
\begin{center}
\includegraphics[width=9.0cm]{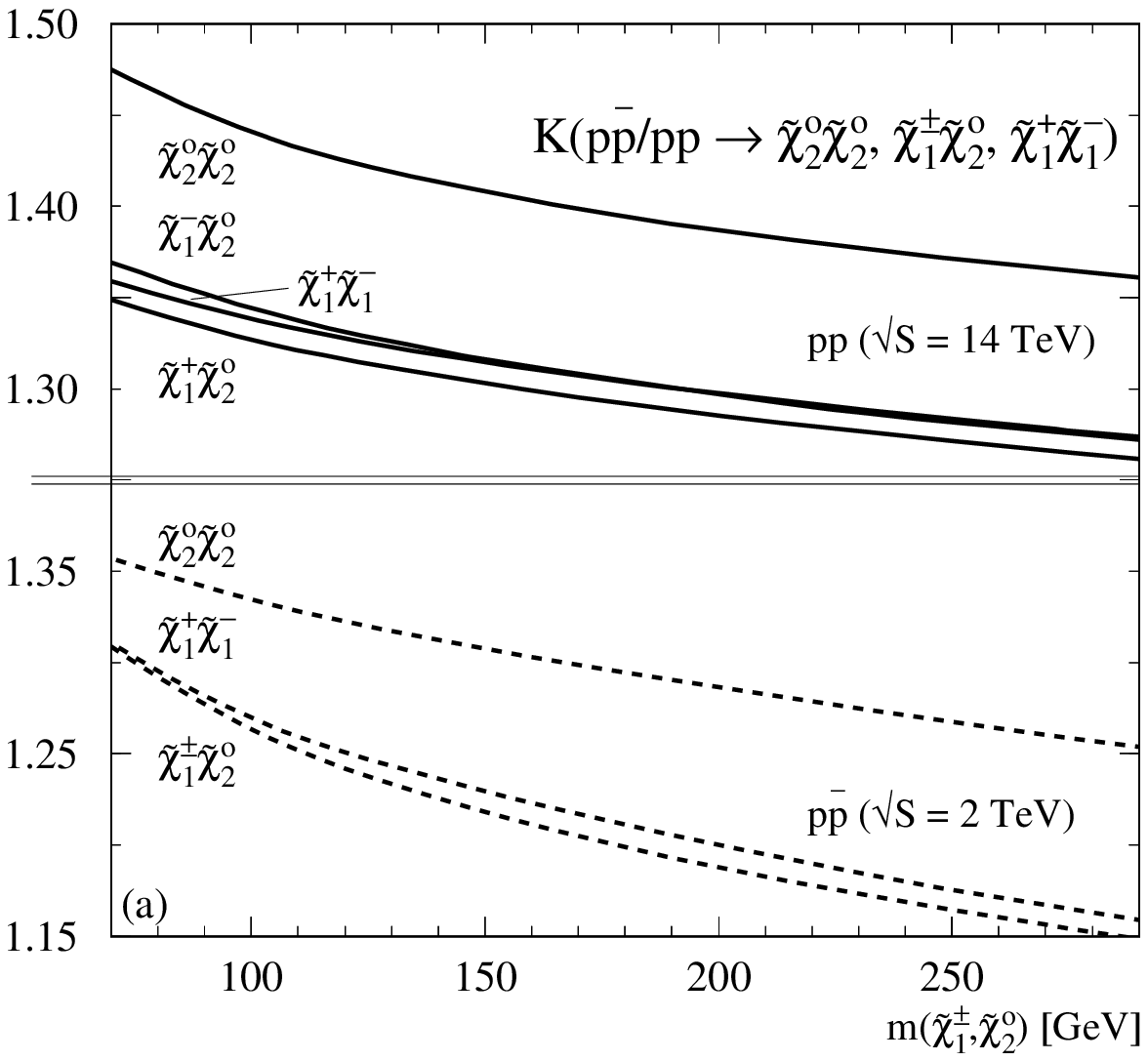} \\[2mm]
\includegraphics[width=9.0cm]{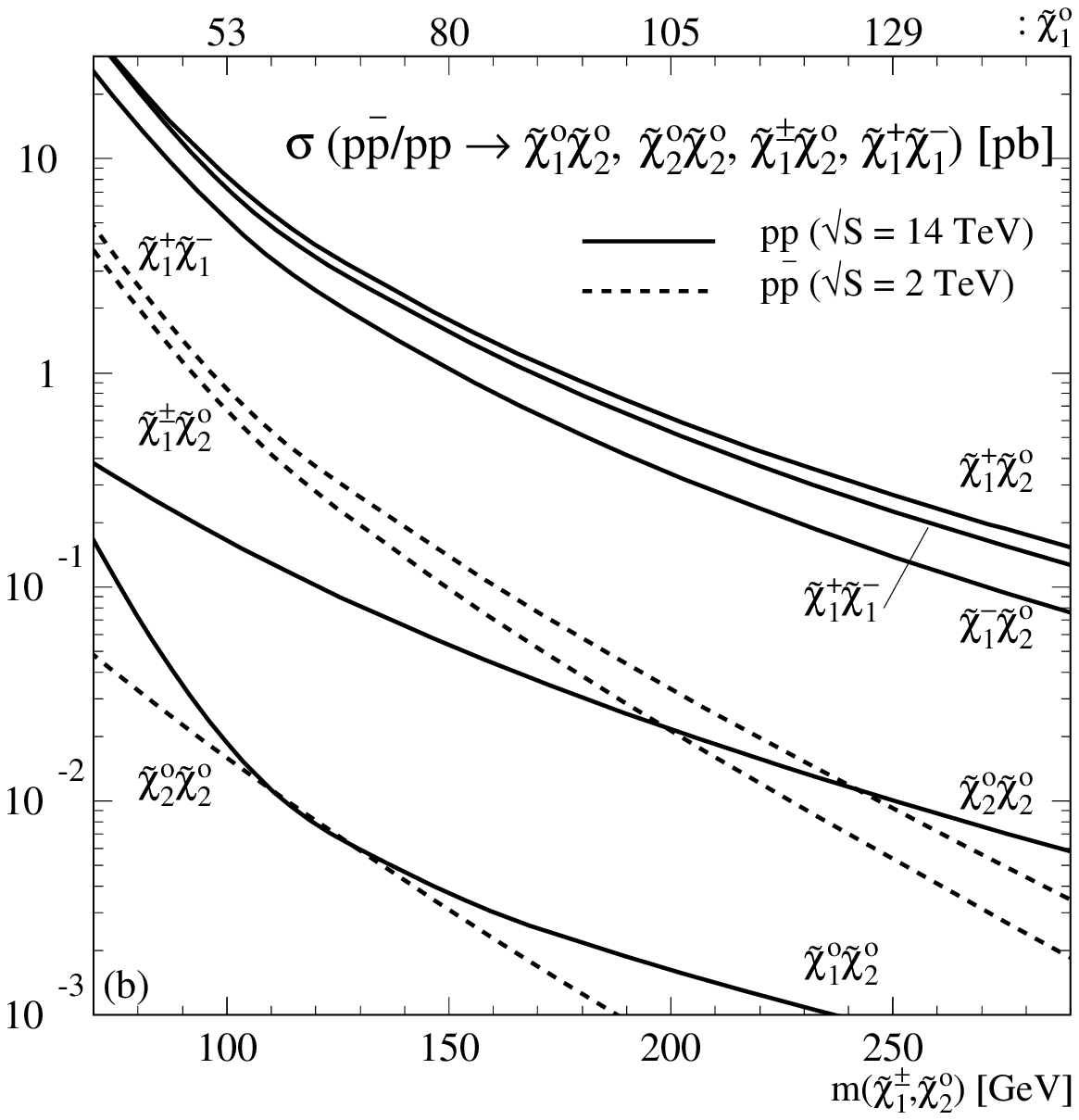} \\[3mm]
\caption[]{\label{fg:cxn_neut} 
    (a) $K$-factors for hadroproduction of chargino/neutralino pairs
    in NLO SUSY QCD, and (b) the NLO cross sections at Tevatron and LHC. The 
    parameters are derived from the mSUGRA point defined in the text,
    but varying the gaugino mass $m_{1/2}$; 
    the factorization/renormalization scale is taken at the average
    chargino/neutralino mass.
    The mass at the lower
    $x$-axis is identified with the chargino/neutralino
    mass or the heavier of the chargino/neutralino 
    masses in the pairs. [The $\cpe$ and $\nnz$ masses nearly coincide.] 
}
\end{center} 
\end{figure}

The theoretical improvement of the predictions for the
chargino/neutralino cross sections is apparent from
Fig.~\ref{fg:scale_neut}. The dependence on the
factorization/renormalization scale $Q$, identified for the sake of
simplicity, is clearly reduced in next-to-leading order QCD with
respect to the factorization-scale dependence in leading order\footnote
{Only for channels with large destructive interferences between 
different exchange amplitudes is the $Q$ dependence not reduced in NLO.
However, in these cases the cross sections are very small and the
channels cannot be explored experimentally.}.  
The cross sections for $p\bar{p}/pp \to \cpe \nnz$ 
are nearly independent of
$Q$. The improvement is most effective for the LHC, since the primary
subprocesses in this case involve sea quarks in addition to valence
quarks. The $K$-factors\footnote{The $K$-factors are defined by
$K=\sigma_{\rm NLO}/\sigma_{\rm LO}$ with all quantities in the cross
sections $\sigma_{\rm NLO}$ and $\sigma_{\rm LO}$ calculated
consistently in next-to-leading order (NLO) and leading order (LO),
respectively.}  of this process, Fig.~\ref{fg:cxn_neut}(a), range from
1.15 to 1.30 at the Tevatron and from 1.25 to 1.35 at the LHC, the
scale $Q$ being fixed, {\it sine restrictione generale}, to the
average final-state mass\footnote{ It is not legitimate to use $Q
=\sqrt{s}$ beyond leading order since this choice of the factorization
scale results in an error of order $\alpha_s$, no matter how
accurately the hard-scattering cross section is
calculated~\cite{soper}.}.  For general chargino/neutralino final
states the $K$-factors extend up to 1.35 and 1.45 at the Tevatron and
the LHC, respectively.

The size of the relevant cross sections for char\-gino/neutralino pair
production at the Tevatron and the LHC is shown in
Fig.~\ref{fg:cxn_neut}(b). The cross sections for other pairs not
shown explicitly in the figures are too small to be accessible
experimentally\footnote{The whole set of cross sections can be
obtained in Fortran code on request; they will be included in ${\tt
PROSPINO}$, which can be accessed at the address
http://www.desy.de/$\sim$spira.}.  The next-to-leading order QCD
corrections increase the cross sections for the production of
chargino/neutralino pairs. The shift in the average of chargino and
neutralino masses that can be probed at the Tevatron and the LHC is
about 15~GeV and 30~GeV, respectively. This corresponds to a
significant improvement, by about $10\%$, of the discovery
limits. Since the chargino/neutralino processes have a colour-flow
similar to Drell--Yan processes, additional effects from higher-order
soft-gluon radiation are expected to be
small~\cite{soft_gluons}.\medskip

\begin{figure}[htb] 
\begin{center}
\includegraphics[width=9.0cm]{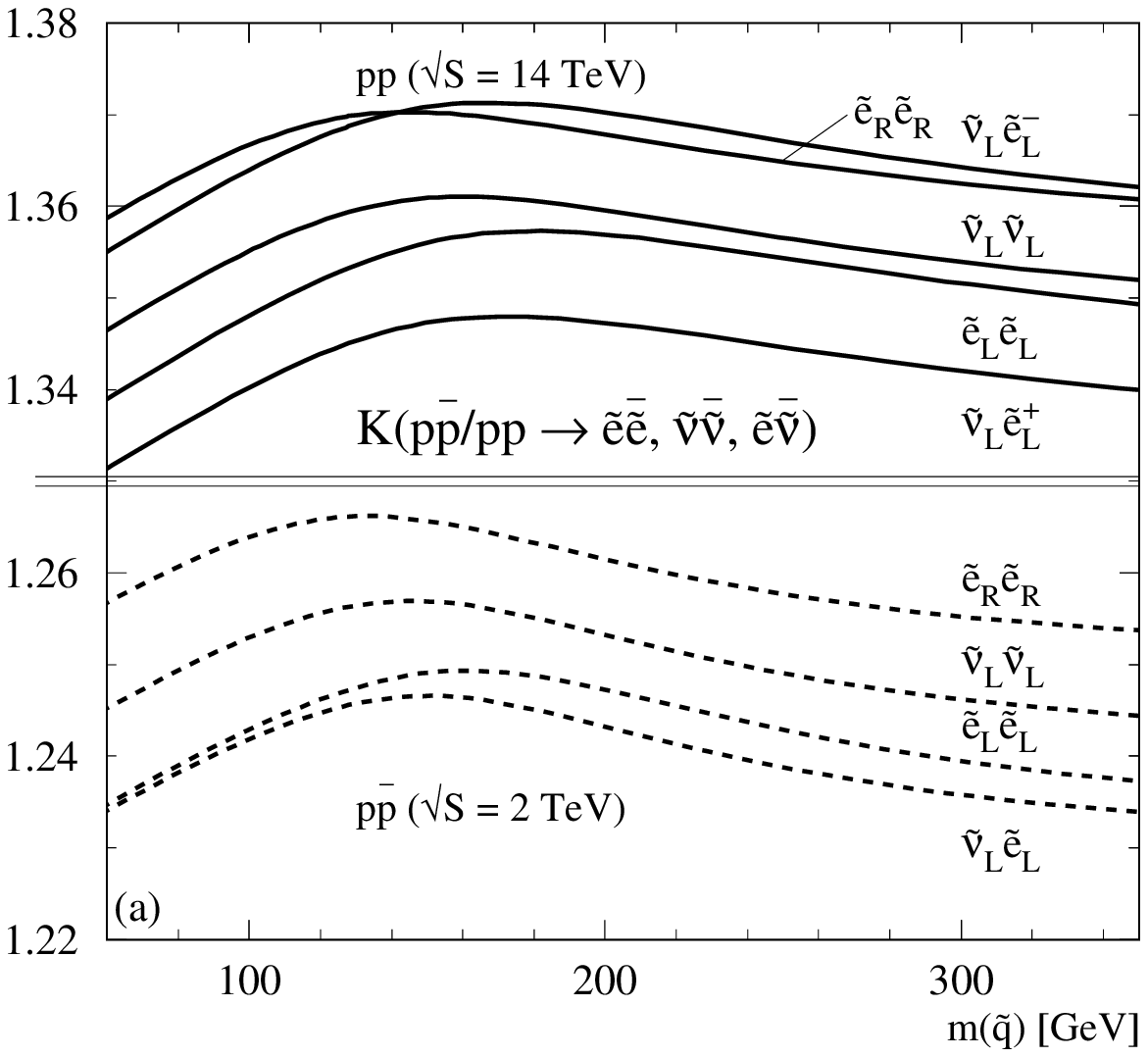} \\[2mm]
\includegraphics[width=9.0cm]{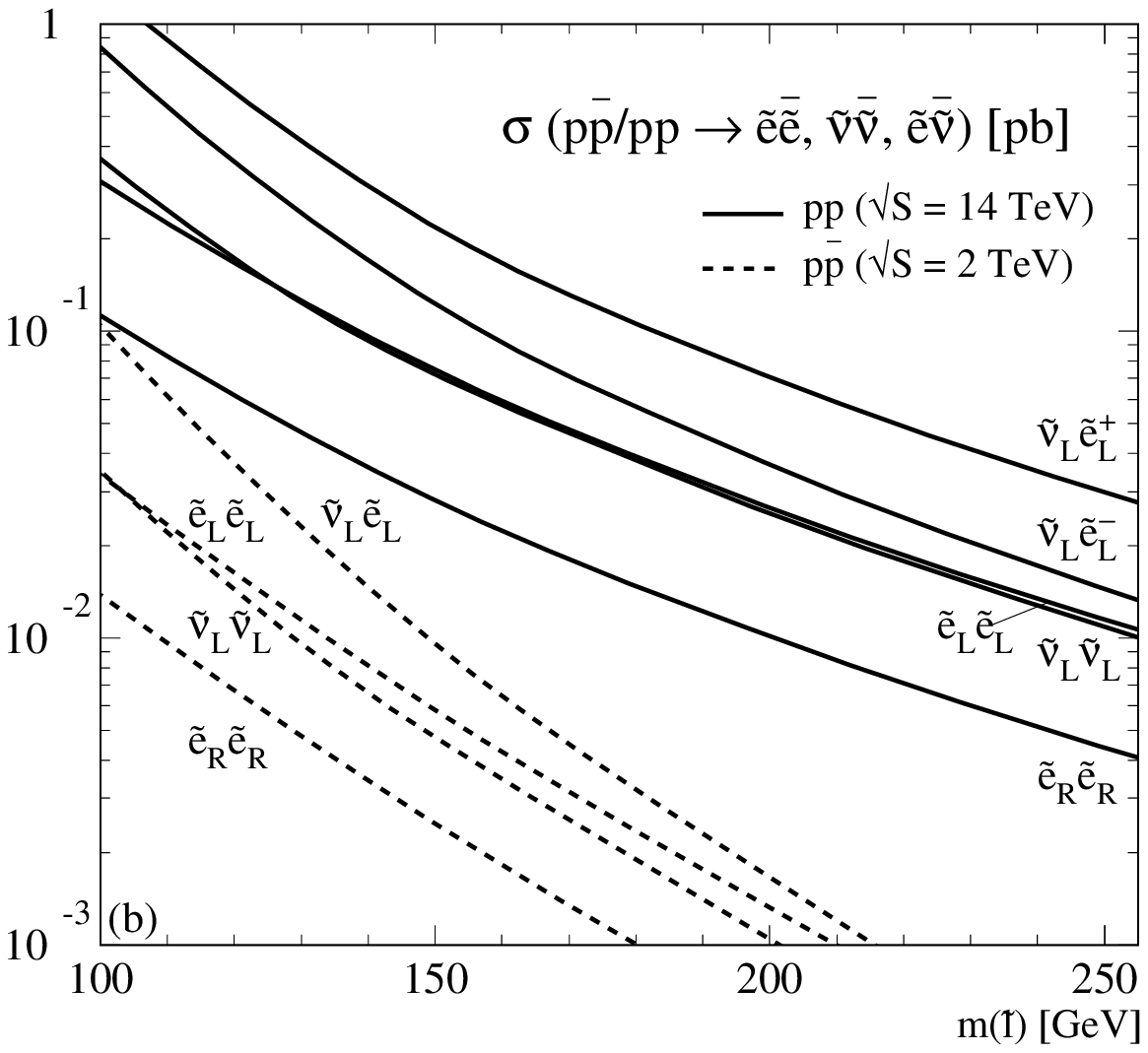} \\[3mm]
\caption[]{\label{fg:slep} 
    (a) $K$-factors for hadroproduction of slepton pairs
    in NLO SUSY QCD as a function of the squark mass $m_{\tilde{q}}$ 
    for a gluino mass of 200~GeV and
    for slepton masses
    ${\tilde{e}_R}/{\tilde{e}_L}/{\tilde{\nu}_L} = 120/150/135$ GeV.
    (b) The NLO cross sections at Tevatron and LHC
    as a function of the slepton masses [${\tilde{e}}$ mass in mixed
    pairs] for squark/gluino masses fixed at 200~GeV. In both figures,
    the factorization/renormalization scale is taken at the
    average slepton mass.    
}
\end{center} 
\end{figure}

\noindent {\bf 2.} Pairs of sleptons, $\sepl \seml, \sepr \semr, 
\snl \snlb$ and $\sepml \snlsnlb$, are generated in
$q\qbqbp$ annihilation by $s$-channel vector-boson exchanges. The QCD
corrections to these Drell--Yan processes have been calculated earlier
in Ref.~\cite{bhr}. In this letter we add the contributions of virtual
squarks and gluinos, completing the analysis consistently to ${\cal
O}(\alpha_s)$. SUSY-QCD corrections affect the $qqV$ vertices, in
analogy to the first column of diagrams in
Fig.~\ref{fg:feyn}(a,b). Since heavy-mass SUSY particles are involved
in the loops, the genuine SUSY corrections are expected to be
considerably smaller than the standard QCD corrections. This is indeed
borne out by detailed calculations, an example of which is presented
in Fig.~\ref{fg:slep}(a). The SUSY-QCD $K$-factors differ little from
the QCD $K$-factors, which are approached in the asymptotic limit of
large $\tilde{q}/\tilde{g}$ masses at the right-hand $y$-axis of the
figure.  A set of typical cross sections for the Tevatron and the LHC
is presented in Fig.~\ref{fg:slep}(b). The cross sections are shown as
a function of the slepton masses for fixed squark/gluino masses.  The
QCD corrections to the production of ${\tilde \mu}$ and ${\tilde
\tau}$ pairs follow the same pattern for equivalent invariant masses
of the pairs.
\medskip

\noindent {\it In Summary:} We have determined the SUSY-QCD
corrections to the cross sections for the production of
chargino/neutralino and slepton pairs at the hadron colliders Tevatron
and LHC consistently to ${\cal O}(\alpha_s)$.  As a result of these
refinements, the theoretical predictions are remarkably stable, being
nearly independent of the renormalization/factorization scales. The
SUSY-QCD corrections are positive, increasing the mass range of
charginos, neutralinos and sleptons that can be covered at these
colliders by as much as ten per cent. This will significantly extend
the area in the supersymmetric parameter space that can be probed at
the Tevatron and LHC beyond the range already accessible at LEP.

%%%%%%%%%%%%%%%%%%%%  ACKNOWLEDGMENTS  %%%%%%%%%%%%%%%%%%%%

\begin{center}
{\bf Acknowledgements}
\end{center}
We are grateful to our experimental CDF and D0 colleagues for helpful
discussions on the experimental chargino/neutralino search programs
during the recent Tevatron Run II Workshop at FNAL . Special thanks go
to T.~Kamon for a very valuable communication in this context.

%%%%%%%%%%%%%%%%%%%%%%%  REFERENCES  %%%%%%%%%%%%%%%%%%%%%%%

\bibliographystyle{plain}

\newpage
%%%%%%%%%%%%%%%%%%%%%%%  FIGURES  %%%%%%%%%%%%%%%%%%%%%%%

\end{document}